\documentclass[twocolumn,preprintnumbers,amsmath,amssymb,prl,longbibliography]{revtex4-1}

\usepackage{graphics}
\usepackage{subfigure}
\usepackage{siunitx}
\usepackage{hyperref}
\hypersetup{colorlinks=true, citecolor=blue, urlcolor=blue, linkcolor=blue}
\allowdisplaybreaks

\begin{document}

\title{Mesoscopic pairing without superconductivity}

\author{Johannes Hofmann}
\email{jbh38@cam.ac.uk}
\affiliation{TCM Group, Cavendish Laboratory, University of Cambridge, Cambridge CB3 0HE, United Kingdom}

\date{\today}

\begin{abstract}
We discuss pairing signatures in mesoscopic nanowires with variable attractive pairing interaction. Depending on wire length, density, and interaction strength, these systems realize a simultaneous bulk-to-mesoscopic and BCS-BEC crossover, which we describe in terms of the parity parameter that quantifies the odd-even energy difference and generalizes the bulk Cooper pair binding energy to mesoscopic systems. We show that the parity parameter can be extracted from recent measurements of conductance oscillations in SrTiO$_3$ nanowires by G. Cheng {\it et al.} [Nature {\bf 521}, 196 (2015)], where it marks the critical magnetic field that separates pair and single-particle currents. Our results place the experiment in the fluctuation-dominated mesoscopic regime on the BCS side of the crossover.
\end{abstract}

\maketitle

In unconventional superconductors, the formation of Cooper pairs and their condensation do not coincide and there are pairing signatures in the normal phase (i.e., above the critical temperature or the critical magnetic field). One picture that captures such pairing without superconductivity is the BCS-BEC crossover, which describes the evolution of a Fermi system from weak to strong attractive pairing interaction. The quantitative distinction between BCS and BEC regimes is the Cooper pair size set by the coherence length $\xi = \hbar v_F/\pi\Delta$ (with $v_F$ the Fermi velocity and $\Delta$ the superconducting gap), which is much larger than the interparticle spacing on the BCS side, $\xi n^{1/d} \gg 1$ ($n$ is the electron density and $d$ the dimension), but decreases rapidly on the BEC side, $\xi n^{1/d} \ll 1$. There,  Cooper pairs are dimers that can form a Bose-Einstein condensate but are also present in the normal state. The BCS-BEC crossover was originally proposed for superconducting semiconductors such as strontium titanate (SrTiO$_3$)~\cite{eagles69,leggett80}, where it is tuned by a change in carrier density~\cite{schooley64,koonce67,lin13,lin14} (with a BCS regime at large doping and a BEC regime at small doping). It has gained a lot of attention over the past decade in applications to ultracold quantum gases~\cite{bloch08,giorgini08,zwerger12,zwerger16}, and may have some bearing on pseudogap physics in high-$T_c$ superconductors~\cite{uemura91,uemura97}. In this Letter, we explore pairing signatures in mesoscopic transport experiments, where in addition to the BCS-BEC crossover, superconductivity can be destroyed by finite level spacing but pairing signatures remain~\cite{anderson59}.
 
\begin{figure}[t]
\subfigure[]{\raisebox{0.45cm}{\scalebox{0.4}{\includegraphics{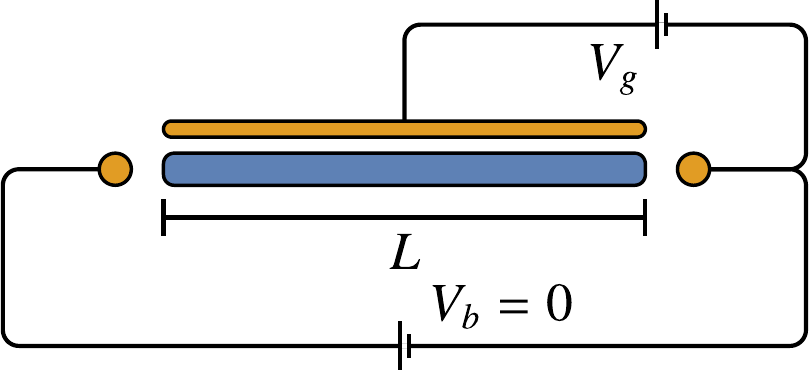}}}\label{fig:1a}}\qquad
\subfigure[]{\scalebox{0.47}{\includegraphics{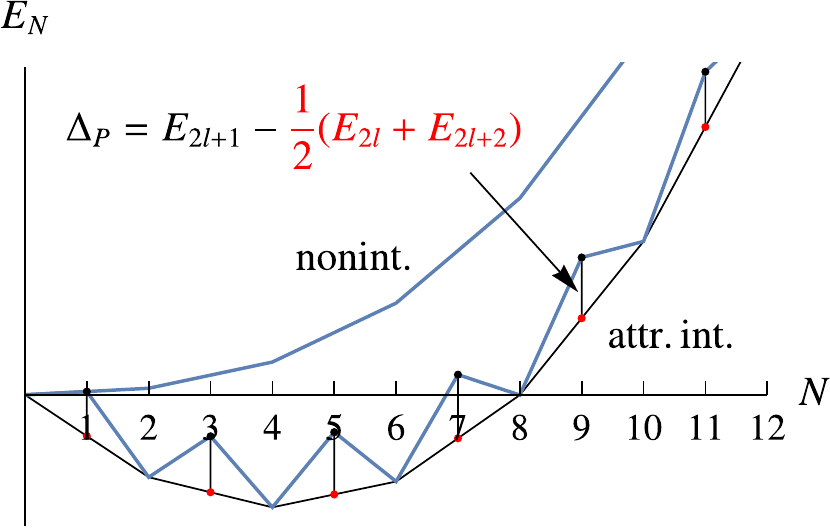}}\label{fig:1b}}
\caption{(a) Schematic of a nanowire single-electron transistor used in Ref.~\cite{cheng15}. (b) Sketch of the ground state energy of a two-component Fermi gas without and with attractive interactions. The ground state energy of an odd-parity state $E_{2l+1}$ (with a single unpaired spin) is larger than the mean of the neighboring even-parity energies. This is the parity effect.}
\end{figure}

The motivation for this work are recent experiments on SrTiO$_3$ nanowires that measure the conductance oscillations of a single-electron SrTiO$_3$ transistor~\cite{cheng15,cheng16} (see Ref.~\cite{pai17} for a review). In Refs.~\cite{cheng15,cheng16}, quantum wires of width $\SI{5}{\nano\meter}$ and length $L=\numrange{0.5}{1} \SI{}{\micro\meter}$ are created at a SrTiO$_3$/LaAlO$_3$ interface and coupled to two leads by tunnel junctions at zero bias as sketched in Fig.~\ref{fig:1a}. The positions of the conductance peaks as a function of gate voltage $V_g$ are independent of an external magnetic field up to $B_p=\SI{2}{\tesla}$, above which the peak Zeeman-splits in two with linear field-dependence~\cite{cheng15}. As $B_p$ is larger by an order of magnitude compared to the critical field of a long wire~\cite{cheng15} or the interface electron gas~\cite{fillistsirakis16}, this is interpreted as direct evidence for pairing without superconductivity as expected for the BEC-side of the BCS-BEC crossover.

%++++++++++++++++++++++++++++++++++++++++
%++++++++++++++++++++++++++++++++++++++++

%++++++++++++++++++++++++++++++++++++++++
\begin{figure*}[t]
\subfigure{\scalebox{0.42}{\includegraphics{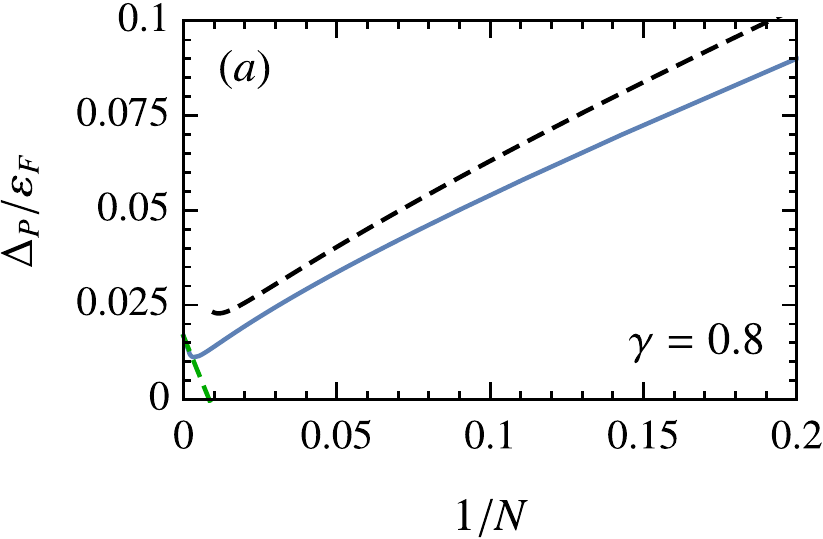}}\label{fig:2a}}
\subfigure{\scalebox{0.41}{\includegraphics{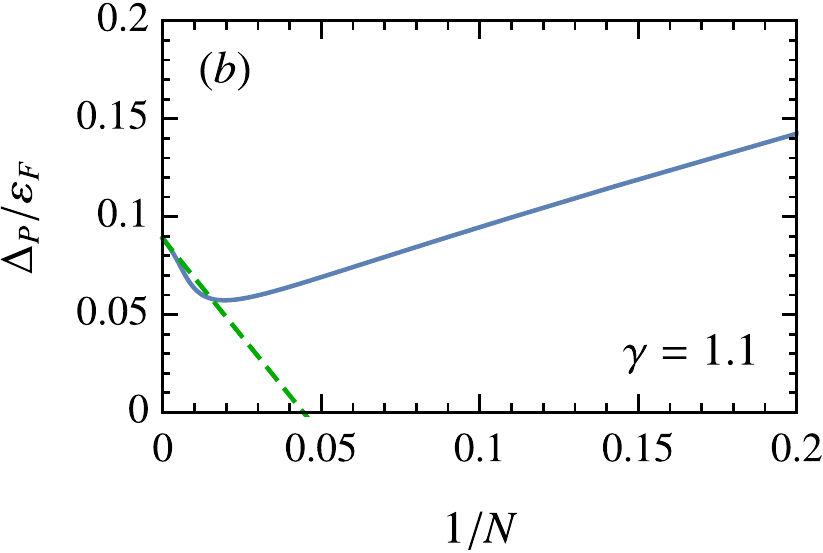}}\label{fig:2b}}
\subfigure{\scalebox{0.4}{\includegraphics{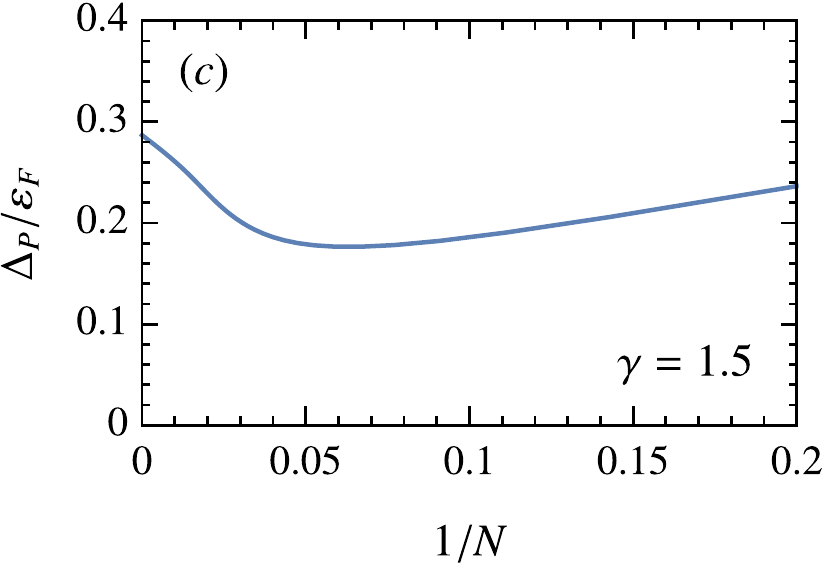}}}
\subfigure{\scalebox{0.4}{\includegraphics{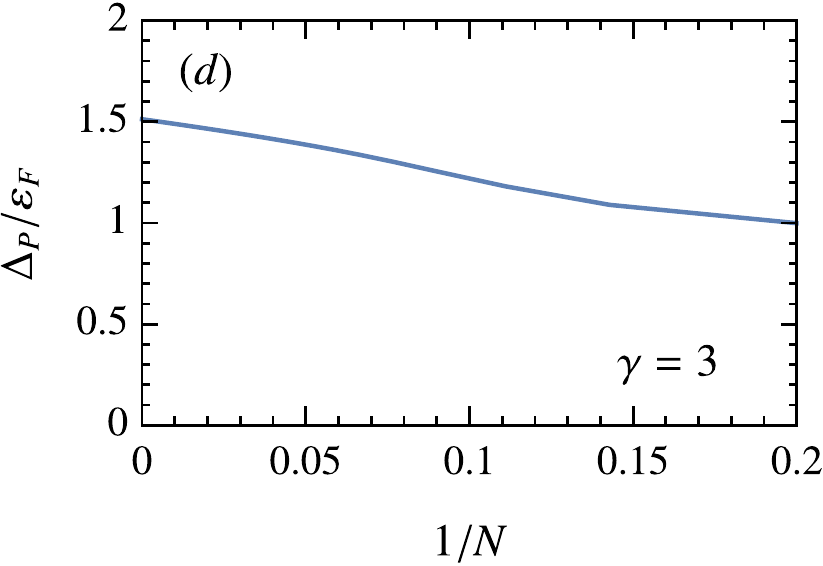}}}
\subfigure{\scalebox{0.4}{\includegraphics{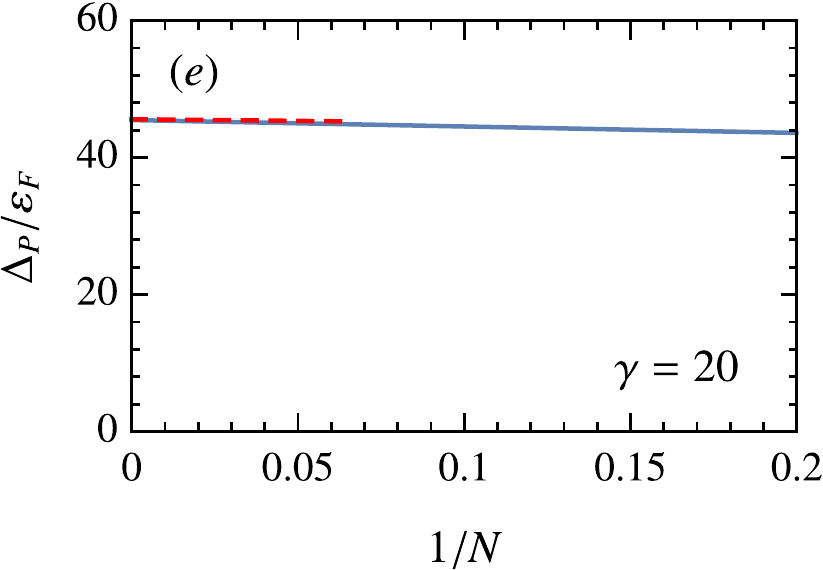}}\label{fig:2e}}
\caption{Parity parameter as a function of particle number $1/N$ for five interaction strengths $\gamma=0.8,1.1,1.5,3,$ and $20$ (a-e). The green and red dashed lines are given in Eq.~\eqref{eq:finitesizecorr}. The black dashed line is the exact BCS result of Ref.~\cite{matveev97}.}
\label{fig:2}
\end{figure*}
%++++++++++++++++++++++++++++++++++++++++
 
In this Letter, we propose an alternative scenario in which the large value of $B_p$ is explained by strong fluctuations in a mesoscopic BCS-type superconductor rather than by the BCS-BEC crossover. We show that the critical field $B_p$ is proportional to the parity parameter~\cite{matveev97}
\begin{align}
\Delta_P &= E_{2l+1} - \frac{1}{2} (E_{2l} + E_{2l+2}) , \label{eq:parity}
\end{align}
a central quantity in the study of mesoscopic superconductors. Here, $E_N$ denotes the ground-state energy of an $N$-electron state in the wire, which (at zero magnetic field) contains an equal number of spin-$\uparrow$ and spin-$\downarrow$ electrons for even total particle number (even parity) and an additional unpaired spin for odd total particle number (odd parity). Generally, in systems with attractive interactions, configurations with even total particle number have an enhanced stability compared to configurations with odd total particle number~\cite{lee07} -- known as the parity effect and illustrated in Fig.~\ref{fig:1b} -- and, hence, the parity parameter is always positive. The parity effect is observed, for example, in nuclei~\cite{lee07}, superconducting nanograins~\cite{tuominen92,lafarge93,tinkham95}, and ultracold Fermi gases~\cite{zuern13,hofmann16}. The parity parameter itself, despite its considerable conceptual importance, has up to date not been observed directly in experiment.

In the context of mesoscopic (BCS) superconductors, the parity parameter was introduced by Matveev and Larkin as a generalization of the bulk BCS gap $\Delta_{\rm BCS}$ to mesoscopic systems~\cite{matveev97}. For BCS-type superconductors, the parity parameter has a minimum if the size of the system $L$ is comparable to the size of a Cooper pair $\xi$, indicating a crossover between a bulk superconducting regime ($\xi/L \ll 1$) and a mesoscopic regime ($\xi/L \gg 1$). In the mesoscopic regime, fluctuations increase the parity parameter as $\Delta_P/\Delta_{\rm BCS} \sim \xi/\Delta_{\rm BCS}\ln (\xi/\Delta_{\rm BCS})$~\cite{matveev97}, which can be much larger than the bulk gap [cf. Figs.~\ref{fig:2a}-\ref{fig:2b}] and which, as we show in this Letter, can account for the large value of $B_p$. Separately, there is also an interaction-induced increase of the parity parameter when moving from the BCS to BEC regime. Our calculations indicate that the first mechanism due to mesoscopic fluctuations plays out in the experiment~\cite{cheng15}.

We begin by introducing a model that describes the simultaneous bulk-to-mesoscopic and BCS-to-BEC crossover in an interacting SrTiO$_3$ nanowire. At small doping in the SrTiO$_3$/LaAlO$_3$ interface, only a single ($d_{xy}$) band is occupied~\cite{khalsa12,sulpizio14}. A wire with length $L$ in this surface layer forms a one-dimensional box with energy levels
\begin{align}
\varepsilon_j = \frac{\hbar^2 \pi^2}{2mL^2} j^2 ,
\end{align}
where $j=1,2,\ldots$ and $m = 0.7 \si{\electronmass}$ is the effective $d_{xy}$ band mass~\cite{santandersyro11}. We include an attractive BCS-type interaction with strength $\lambda$ that couples pairs of time-reversed states with opposite spin and consider the Hamiltonian
\begin{align}
\hat{H} &= \sum_{j\sigma} (\varepsilon_j + \sigma h) \hat{c}_{j\sigma}^\dagger \hat{c}_{j\sigma}^{} - \lambda \sum_{ij} \hat{c}_{i\uparrow}^\dagger \hat{c}_{i\downarrow}^\dagger \hat{c}_{j\downarrow}^{} \hat{c}_{j\uparrow}^{} , \label{eq:model}
\end{align}
where $\hat{c}_{j\sigma}^\dagger$ creates an electron in state $j$ with energy $\varepsilon_j$ and spin $\sigma = \pm1$ (or $\sigma=\uparrow,\downarrow$). An external magnetic field $h = \frac{1}{2} g \mu_B B$ shifts the single-particle energies through the Zeeman term, where $g \approx 1.2$ is the Land\'{e} factor~\cite{cheng15} and $\mu_B$ the Bohr magneton. A dimensionless interaction strength $\gamma$ is set by the ratio of interaction energy $E_{\rm int} = \frac{\lambda}{4} L^2 n^2$ and kinetic energy $E_{\rm kin} = \frac{\pi^2}{12} \frac{\hbar^2}{m} L n^3$ as
\begin{align}
\gamma &= \frac{m \lambda L}{\hbar^2 n} .
\end{align}
The large and small doping regime corresponds to the BCS and BEC limit (with weak and strong attractive interaction, respectively).

The model~\eqref{eq:model} is an example of a Richardson-Gaudin model, which can be solved exactly for arbitrary particle number~\cite{richardson63,richardson64,richardson65,richardson66,dukelsky04}. Since the interacting ground and excited states evolve adiabatically from their noninteracting counterparts, an interacting configuration is characterized by the occupation of states in a box, i.e., by a set $\{J_1,\ldots,J_{n_p}\}$ of $n_p$ levels occupied by pairs and a set ${\cal B} = \{ \{k_1, \sigma_1\}, \ldots, \{k_{n_s}, \sigma_{n_s}\}\}$ of $n_s$ levels occupied by single electrons. ${\cal B}$ is called the set of blocked levels because the interaction term in Eq.~\eqref{eq:model} only couples empty or doubly-occupied states, i.e., singly-occupied states do not participate in the interaction. The $N=2n_p+n_s$-particle wavefunction of this state is~\cite{vondelft00,dukelsky04}
\begin{align}
|\Psi_N\rangle &= \biggl(\prod_{\{\ell,\sigma\}\in {\cal B}} \hat{c}_{\ell\sigma}^\dagger\biggr) \biggl(\prod_{\mu=0}^{n_p} \hat{B}_{J_\mu}^\dagger\biggr) |0\rangle ,
\end{align}
where $ |0\rangle$ is the vacuum state without particles and
\begin{align}
\hat{B}_J^\dagger &= \lambda C_J \sum_{j\not\in {\cal B}} \frac{\hat{c}_{j\uparrow}^\dagger \hat{c}_{j\downarrow}^\dagger}{2 \varepsilon_{j} - E_J}
\end{align}
with normalization constant $1/(\lambda C_J)^2 = \sum_{j\not\in {\cal B}} 1/|2 \varepsilon_{j} - E_J|^2$. The wavefunction is determined by the set of roots $\{E_J\}$ that solve the Richardson equations~\cite{richardson63,vondelft00,dukelsky04}
\begin{align}
\frac{1}{\lambda} - \sum_{j\not\in {\cal B}}\frac{1}{2 \varepsilon_{j} - E_{J_\nu}} + \sum_{\mu=1, \mu\neq\nu}^{n_p} \frac{2}{E_{J_\mu} - E_{J_\nu}} &= 0 , \label{eq:richardson}
\end{align}
which can be solved numerically~\cite{richardson66} much more efficiently than by exact diagonalization of the model~\eqref{eq:model}. The energy of the state $|\Psi_N\rangle$ is given by~\cite{richardson63}
\begin{align}
E_N = \sum_{i=1}^{n_p} E_{J_i} + \sum_{\{\ell,\sigma\}\in {\cal B}} (\varepsilon_{\ell} +\sigma h) .\label{eq:gsenergy}
\end{align}
An important point is that the roots $E_J$ are independent of the field $h$, which affects the ground state energy $E_N$ only through the Zeeman shift of unpaired electrons in blocked levels.

We first discuss the behavior of the parity parameter~\eqref{eq:parity} along the BCS-BEC crossover. To this end, we solve the Richardson equations~\eqref{eq:richardson} numerically for up to $2000$ particles in an interaction range $0 \leq \lambda/\varepsilon_1 \leq 2500$. In order to compare few- and many-body regimes with similar pairing behavior, we determine the parity parameter as a function of total particle number while keeping both the total density $n=N/L$ and the dimensionless interaction strength~$\gamma$ fixed. Figure~\ref{fig:2} shows the result of this calculation as a function of inverse particle number for five interaction strengths $\gamma = 0.8,1.1,1.5,3,$ and $20$. Two main results are directly apparent from the figure: first, along the BEC-BCS crossover, the parity parameter increases strongly with increasing interaction strength $\gamma$. In the weak-coupling and strong-coupling limit, the explicit form at large $N$ is
\begin{align}
\frac{\Delta_P}{\varepsilon_F} &= \begin{cases}
8 \, e^{-\tfrac{\pi^2}{2\gamma}} - \dfrac{2}{N} & \gamma \to 0+ \\[1ex]
\dfrac{\gamma (\gamma+2)}{\pi^2} - \dfrac{8 \gamma}{\pi^2 N} & \gamma \to \infty
\end{cases} , \label{eq:finitesizecorr}
\end{align}
which are shown as a green dashed line and a red dashed line in Figs.~\ref{fig:2a},~\ref{fig:2b}, and~\ref{fig:2e}, respectively. The first term is the bulk mean field result and the second term is a finite-size correction. The bulk limit at weak coupling corresponds to the mean field gap $\Delta_{\rm BCS} = 8 \varepsilon_F e^{-\pi^2/2\gamma}$, which is the energy cost to add an additional unpaired fermion to a condensate of Cooper pairs. On the BEC side, however, the bulk parity parameter is equal to the absolute value of the chemical potential $\mu_{\rm BEC} = - \gamma^2/\pi^2$, which corresponds to the bound state energy of a dimer: The parity parameter generalizes to a pair breaking energy along the crossover. Second, on the BCS side for $\gamma \lesssim 3$, there is a clear separation between a bulk and a mesoscopic regime as the particle number is lowered (marked by a minimum in the parity parameter for $\xi \sim L$), and the parity parameter can be much larger than its bulk value [Figs.~\ref{fig:2a} and~\ref{fig:2b}]. With increasing interaction strength, however, this minimum is pushed to smaller particle number and eventually disappears on the BEC side -- the size of the dimer is so small that no finite-size effects can be detected even for the shortest wires.

Note that the weak-coupling limit can be connected to models of mesoscopic pairing for metallic nanograins~\cite{matveev97,vondelft01,dukelsky00,sierra00,yuzbashyan05}, where the total particle number is large and pairing is restricted to the vicinity of the Fermi surface. Here, an attractive interaction of the type used in Eq.~\eqref{eq:model} is combined with equally spaced single-particle energy levels $\varepsilon_j = j \delta \varepsilon$ with $j = 0, \pm 1, \ldots \pm l_{\rm Debye}$, where $\omega_{\rm D} = l_{\rm Debye} \delta \varepsilon$ is the Debye frequency. The crossover is studied as a function of $\delta \varepsilon/\Delta_{\rm BCS}$ (or $\delta \varepsilon/\omega_{\rm D}$) while keeping the system at half-filling, interpolating between a bulk limit for $\delta \varepsilon \gg \Delta_{\rm BCS}$ and a fluctuation-dominated limit for $\delta \varepsilon \ll \Delta_{\rm BCS}$. Using for our model $\delta \varepsilon = \frac{\hbar^2 k_F}{m} \frac{\pi}{L}$ in Eq.~\eqref{eq:finitesizecorr}, we have for the first correction in the bulk limit $\Delta_P = \Delta_{\rm BCS} - \frac{\delta \varepsilon}{2}$, in agreement with the result for nanograins~\cite{matveev97,yuzbashyan05}. In the mesoscopic BCS-limit, the enhancement of the parity parameter is $\Delta_P = \delta \varepsilon/2\ln(\delta\varepsilon/\Delta_{\rm BCS}) \gg \Delta_{\rm BCS}$~\cite{matveev97}. This expression is shown as a black dashed line in Fig.~\ref{fig:2a} and agrees well with our results for smaller $\gamma$.

%++++++++++++++++++++++++++++++++++++++++
\begin{figure}[t]
\subfigure[]{\scalebox{0.5}{\includegraphics{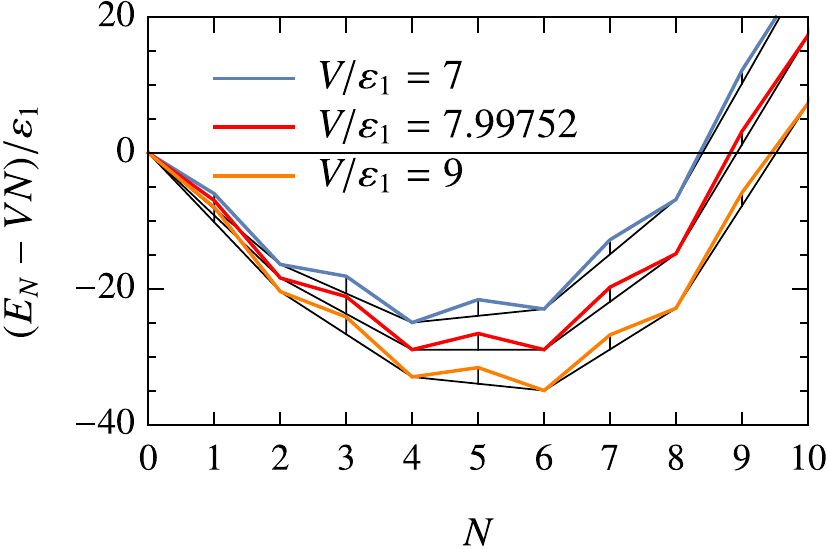}}\label{fig:3a}}
\subfigure[]{\scalebox{0.475}{\includegraphics{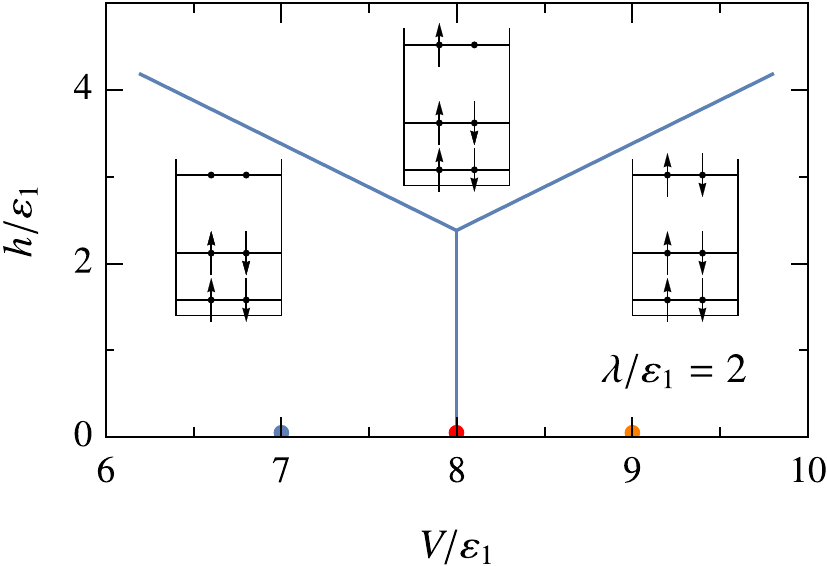}}\label{fig:3b}}
\caption{(a) Energy $E_N - VN$ in the quantum wire as a function of particle number with $\lambda/\varepsilon_1 = 2$ for three different electrostatic potentials $V/\varepsilon_1 = 7,7.99752$, and $9$. The minimum marks the ground state. (b) Ground state phase diagram in the $V$-$h$ plane.}
\end{figure}
%++++++++++++++++++++++++++++++++++++++++

We now turn our attention to the experiment~\cite{cheng15} and show that the parity parameter defines the critical magnetic field $B_p$ that marks the bifurcation of the conductance peak. For the quantum wire in contact with the leads and side gate, the ground state configuration at zero temperature is obtained by minimizing~\cite{matveev97}
\begin{align}
\Omega(V) &= \min_N (E_N - V N) , \label{eq:Omega}
\end{align}
where $V$ is the electrostatic potential in the wire which depends linearly on the gate voltage. The charging energy $E_C = e^2/C_\Sigma$ is negligible because of the large dielectric constant of SrTiO$_3$~\cite{mueller79}. Figure~\ref{fig:3a} illustrates the zero magnetic field case using the exact solution of the Richardson model~\eqref{eq:model} with $\lambda/\varepsilon_1 = 2$ for three different values of the potential $V/\varepsilon_1 = 7,7.99752$, and $9$. The thin black lines interpolate between the even-parity ground state energies. As in Fig.~\ref{fig:1b}, the excess energy of the odd-parity states (marked by the vertical black lines) is the parity parameter~\eqref{eq:parity}. A unique minimum of Eq.~\eqref{eq:Omega} determines the ground state particle number. In this case, the wire is in a Coulomb blockade regime provided that $k_B T < \Delta_P$. Importantly, because of the parity effect, this ground state has even total particle number. In the special case where $V$ is fine-tuned such that Eq.~\eqref{eq:Omega} has two minima, pairs can tunnel on and off the two leads kept at zero bias, and a pair current will flow. This corresponds to the condition that the electrostatic potential $V$ is tuned equal to the pair chemical potential: $V_c(N) = (E_{N+2} - E_N)/2$. Since the ground state energy of an even-parity state does not depend on the external field, $E_{2l}(h) = E_{2l}(h=0)$, this picture is unchanged for weak fields. However, an odd-parity state lowers its ground state energy in a magnetic field through the Zeeman shift of the unpaired electron [cf. Eq.~\eqref{eq:gsenergy}]: $E_{2l+1}(h) = E_{2l+1}(h=0) - h$. This defines a critical field $h_p$ above which the even-parity state is no longer the ground state. Above this field, a single-particle tunneling current can flow. This is summarized in Fig.~\ref{fig:3b}, which shows the ground state phase diagram of the wire in the $V$-$h$ plane with the same parameter values as in Fig.~\ref{fig:3a}. The vertical line denotes the pair current, and the slanted lines denote the single-particle current. This is precisely the form observed in the experiment~\cite{cheng15}, where a similar qualitative picture was obtained from a numerical solution of a sixteen-site Hubbard chain. From Eq.~\eqref{eq:gsenergy} (and Fig.~\ref{fig:3a}), we see that this critical field $h_p$ is given by the parity parameter:
\begin{align}
h_p = \frac{1}{2} g \mu_B B_p = \Delta_P .
\end{align}
This is a central result of this work.

Having gained a microscopic understanding of the critical field $B_p$ in terms of the parity parameter, we can fit our results to the experiment~\cite{cheng15} to obtain an interaction strength $g$ and particle number $N$. The electrostatic potential in the wire is related to the gate voltage as $V = e \alpha V_g$ with a lever arm $\alpha \approx 0.1$~\cite{cheng15} (we neglect the contribution of background charge and lead potentials as they only shift the position of the conductance peaks). The critical field at which the conductance peaks split is $B_p \approx \SI{1.8}{\tesla}$~\cite{cheng15}, corresponding to an energy $h_p = \frac{1}{2} g \mu_B B_p \approx \SI{63}{\micro\electronvolt}$. In addition, the separation $\Delta V_g(N)$ of two adjacent conductance peaks bordering a Coulomb blockade region with $N$ particles is $\Delta V_g(N) = - \frac{1}{e\alpha} \Delta_P^{(2)}$, where $\Delta_P^{(2)}$ is a pair parity parameter,
\begin{align}
\Delta_P^{(2)} &= E_N - \frac{1}{2} (E_{N-2} + E_{N+2}) ,
\end{align}
which only depends on even-parity states with $\Delta_P^{(2)} = - (N+1) \varepsilon_1$ for weak interactions. Experimentally, $\Delta V_g \approx \SI{4.5}{\milli\volt}$ from Fig. 2(a) of Ref.~\cite{cheng15}. A simultaneous fit of our calculations to parity parameter and peak splitting gives $N= \numrange{800}{900}$ with $\lambda/\varepsilon_1 = \numrange{65}{75}$ or $\gamma \approx \numrange{0.4}{0.5}$, respectively. This particle number corresponds to a typical two-dimensional density $n_{\rm 2d}=\numrange{1.6}{1.8e13}\SI{}{\cm^{-2}}$. The smallness of the coupling constant $\gamma\approx \numrange{0.4}{0.5}$ puts the system on the BCS side of the crossover, and with $\xi/L \approx \numrange{2}{7}$ (or $\delta \varepsilon/\Delta_{\rm BCS} \approx \numrange{20}{70}$), the wire is in the mesoscopic limit, where the parity parameter exceeds the bulk gap [cf. Fig.~\ref{fig:2a}], $\Delta_P/\Delta_{\rm BCS} \approx \numrange{3}{8}$ here. Assuming that for a given gate voltage, the particle density $n$ and effective interaction strength $\gamma$ do not change with the length of the wire, the observed order-of-magnitude enhancement of $B_p$ compared to $B_{c2}$ in Ref.~\cite{cheng15} can thus be accounted for by a strong fluctuation-induced enhancement of the parity parameter in the mesoscopic regime, which is characteristic of the BCS rather than the BEC regime [cf. Fig.~\ref{fig:2}].

\begin{figure}[t]
\subfigure[]{\scalebox{0.5}{\includegraphics{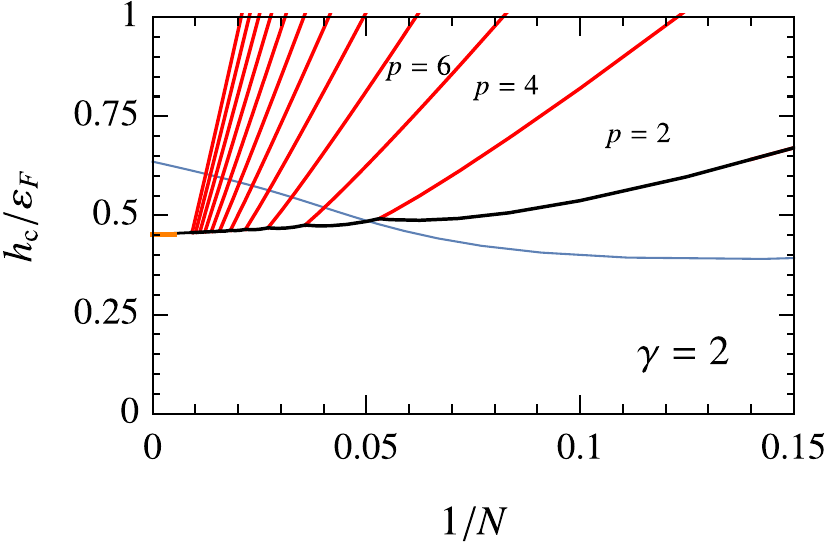}}\label{fig:4a}}\quad
\subfigure[]{\scalebox{0.473}{\includegraphics{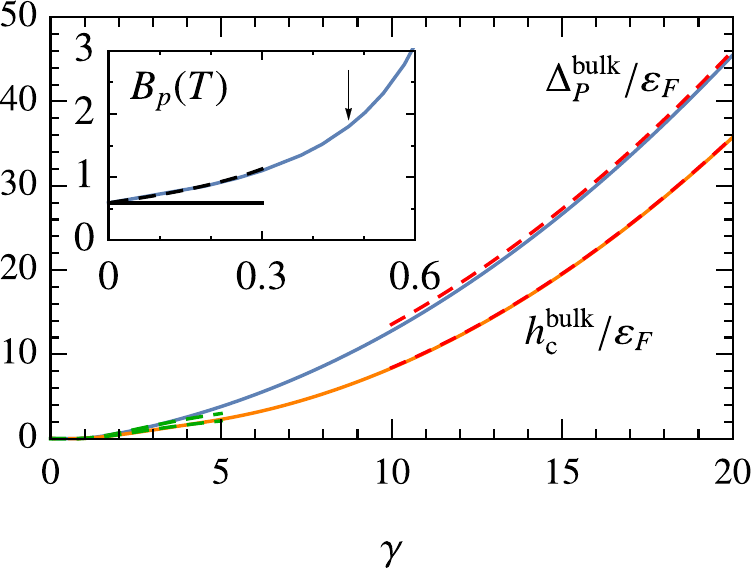}}\label{fig:4b}}
\caption{(a) Ground state phase diagram of the even-parity system in a magnetic field with $\gamma=2$. (b) Parity parameter and critical Clogston field in the bulk limit $N\to \infty$ as a function of interaction strength $\gamma$. Dashed lines indicate the limits of Eq.~\eqref{eq:finitesizecorr}. Inset: critical field $B_p$ as a function of particle number for fixed $\lambda/\varepsilon_1 = 65$. Black lines correspond to the asymptotic limits.}
\end{figure}

We briefly compare the parity parameter with the critical Clogston magnetic field $h_c$ above which the ground state of an even-parity state has finite polarization $p = N_\uparrow - N_\downarrow$. There is a fundamental distinction in these scales as the Clogston field compares the energies of equal-parity states, whereas the parity parameter compares energies with opposite parity. Figure~\ref{fig:4a} shows the ground state phase diagram for fixed parity as a function of particle number and magnetic field at $\gamma=2$. The black line is the Clogston field and red lines indicate transitions between ground states with different polarization. For finite $N$, the polarized state contains pairs (i.e., $p<N$) and is thus interacting, whereas in the bulk limit of large $N$ and fixed density, the Clogston field marks the transition to a noninteracting partially polarized state, as in the BCS case~\cite{braun99}. For comparison, we plot the parity parameter (blue line), which is larger than the Clogston field above a critical particle number, implying that in this regime the bifurcation of the Coulomb conductance peak will be preempted by a transition to an equal-parity polarized state. Figure~\ref{fig:4b} compares the bulk values of parity parameter and Clogston field. Both quantities are of comparable magnitude along the BEC-BCS crossover, and their ratio interpolates between the standard BCS value $h_{\rm c}/\Delta = 1/\sqrt{2}$ and $h_{\rm c}/\Delta = 1$ in the BEC limit.

%++++++++++++++++++++++++++++++++++++++++
%++++++++++++++++++++++++++++++++++++++++

To further illustrate the strong enhancement of the parity parameter in the mesoscopic BCS regime, we show in the inset of Fig.~\ref{fig:4b} the critical field for the fitted parameter values for different particle number, where the black arrow indicates the previous fit with $N=800$. At small $\gamma$, the parity parameter saturates to a constant nonzero value $\Delta_P = \frac{\lambda}{2}$ given by perturbation theory (continuous black line), where the dashed line indicates the logarithmic Matveev-Larkin correction~\cite{matveev97}. Since the critical field $B_c$ vanishes exponentially at small $\gamma$, there is indeed an infinite enhancement of the ratio $B_p/B_c$ as $\gamma\to0$. A similar effect does not exist in the BEC limit.

In summary, we discuss superconductors in the simultaneous crossover between the bulk and mesoscopic regime as well as BEC-type and BCS-type superconductivity. The central quantity that parameterizes the crossover is the Matveev-Larkin parity parameter [Eq.~\eqref{eq:parity}], which is enhanced by mesoscopic fluctuations on the BCS side of the crossover or by an increased interaction on the BEC side. We point out that the parity parameter can be observed in recent experiments on SrTiO$_3$ nanostructures~\cite{cheng15}, where it is linked to the critical magnetic field that marks the transition between pair-tunneling current and single-particle current. Our calculations place the experiment~\cite{cheng15} on the BCS and the mesoscopic side of the crossover. Given the versatility of current experiments, this could allow a systematic experimental investigation of mesoscopic superconductors.

\begin{acknowledgments}
I would like to thank Antonio Garc\'{i}a-Garc\'{i}a and Alejandro Lobos for helpful discussions. This work is supported by Gonville and Caius College, Cambridge.
\end{acknowledgments}

\bibliography{bib}

\end{document}